\documentclass[showpacs,aps,prl,reprint,superscriptaddress,preprintnumbers,nolinenumbers]{revtex4-1}

\usepackage{amsmath}
\usepackage{amssymb}
\usepackage{graphicx}
\usepackage{dcolumn}
\usepackage[dvipsnames]{xcolor}
\usepackage{amsthm,amssymb}
\usepackage{mathrsfs}
\usepackage{color}
\usepackage{endnotes}
\usepackage{bm}
\usepackage{epsfig}
\usepackage{array}
\usepackage{ulem}
\usepackage{soul}

\newcommand{\etal}{{\it et al.}}
\newcommand{\ie}{{\it i.e.\ }}

\newcommand{\parallelsum}{\mathbin{\!/\mkern-5mu/\!}}

\begin{document}



\title{
Magnetotransport evidence of a potential low-lying Dirac node in NbAl$_3$
}

\author{Ying~Kit~Tsui}
\affiliation{Department of Physics, The Chinese University of Hong Kong, Shatin, Hong Kong, China}
\author{Chia-Nung~Kuo}
\affiliation{Department of Physics, National Cheng Kung University, Tainan 70101, Taiwan}
\affiliation{Taiwan Consortium of Emergent Crystalline Materials, National Science and Technology Council, Taipei 10601, Taiwan}
\author{Makoto Shimizu}
\affiliation{Department of Physics, Graduate School of Science, Kyoto University, Kyoto 606-8502, Japan}
\author{Yajian~Hu}
\affiliation{Department of Physics, The Chinese University of Hong Kong, Shatin, Hong Kong, China}
\author{Youichi Yanase}
\affiliation{Department of Physics, Graduate School of Science, Kyoto University, Kyoto 606-8502, Japan}
\author{Chin~Shan~Lue}
\affiliation{Department of Physics, National Cheng Kung University, Tainan 70101, Taiwan}
\affiliation{Taiwan Consortium of Emergent Crystalline Materials, National Science and Technology Council, Taipei 10601, Taiwan}
\author{Wei~Zhang}
\email[]{wzhang@phy.cuhk.edu.hk}
\author{Swee~K.~Goh}
\email[]{skgoh@cuhk.edu.hk}
\affiliation{Department of Physics, The Chinese University of Hong Kong, Shatin, Hong Kong, China}

\date{\today}

\begin{abstract}
NbAl$_3$ is a novel semimetal with a type-II Dirac node $\sim$230~meV above the Fermi energy. We have performed both out-of-plane ($B\parallelsum c$) and in-plane magnetotransport measurements ($B\perp c$) on single-crystalline NbAl$_3$. In our out-of-plane data, we observe an interesting linear component in the transverse magnetoresistance, and the mobility spectrum analysis of the out-of-plane data reveals an emergence of high-mobility electrons at low temperatures. Near $B\parallelsum c$, Shubnikov-de Haas oscillations are discerned in the magnetoresistance. The oscillation frequencies agree with the density functional theory calculation, the same theory that shows that the Dirac node is far above the Fermi energy. Therefore, the out-of-plane results cannot be attributed to the type-II Dirac node but suggest NbAl$_3$ has additional Dirac or Weyl nodes close to the Fermi energy. To support this, we examine the in-plane data obtained with the magnetic field perpendicular to the tilting direction of the type-II Dirac cone. Such field direction excludes the possibility of chiral anomaly from the predicted type-II Dirac node. Remarkably, we observe the planar Hall effect, anisotropic magnetoresistance, and negative longitudinal magnetoresistance. These in-plane results are a strong indication of chiral anomaly unrelated to the previously established type-II Dirac node, pointing to the presence of additional Dirac or Weyl nodes near the Fermi energy. Our new density functional theory calculation reveals a type-I Dirac node $\sim$50~meV below the Fermi energy that has previously been overlooked. We argue that the exotic transport phenomena observed in NbAl$_3$ can be attributed to the newly identified type-I Dirac node.
\end{abstract}

\maketitle

Topological semimetals such as the Dirac and Weyl semimetals host exotic relativistic quasi-particles and interesting electronic band structures~\cite{Gorbar2021Book, Yan2017Topo, Armitage2018Topo, Hu2019Topo, Lv2021Topo, Soluyanov2015WeylII, Chang2017MA3}. While the type-I Dirac or Weyl semimetal has an upright or slightly tilted Dirac or Weyl cone, the type-II counterpart has a strongly tilted Dirac or Weyl cone that intersects the Fermi energy, resulting in the Dirac or Weyl point living at the boundary between an electron pocket and a hole pocket~\cite{Soluyanov2015WeylII, Chang2017MA3}. Such dispersion relation violates Lorentz symmetry~\cite{Soluyanov2015WeylII, Chang2017MA3, Yan2017Topo, Armitage2018Topo, Lv2021Topo} and has no equivalence in high energy physics. Therefore, the type-II Dirac and Weyl semimetal provide a playground of novel quasi-particles.

One exciting aspect of Dirac and Weyl semimetals is the realization of chiral anomaly in condensed matter system~\cite{Nielsen1983Chiral, Son2013Nmr, Burkov2015Chiral, Udagawa2016Chiral, Nandy2017Phe, Burkov2017Phe}. A consequence of chiral anomaly is the observation of planar Hall effect~(PHE) and anisotropic magnetoresistance~(AMR)~\cite{Nandy2017Phe, Burkov2017Phe, Zhong2023Phe}. For PHE and AMR, the exerted current, magnetic field, and electric field (transverse or longitudinal) all lie in the same plane. In Dirac or Weyl semimetals, under a magnetic field, Landau quantization occurs and the lowest Landau level corresponding to each Weyl cone is chiral~(one right-moving parellel to the magnetic field and one left-moving antiparallel to the field)~\cite{Nielsen1983Chiral, Yan2017Topo, Liang2018Phe, Armitage2018Topo, Hu2019Topo, Lv2021Topo, Zhong2023Phe}. Without an applied electric field, the number of right movers and left movers are equal. However, upon applying the electric field parallel to the magnetic field, there is an imbalance between the right and left movers. This constitutes a chiral current along the direction of the magnetic field. Thus, the resistivity decreases with increasing magnetic field when the field is parallel to the current, \ie negative longitudinal magnetoresistance (MR)~\cite{Son2013Nmr,Burkov2014Chiral,Burkov2015Chiral}, but it increases when the in-plane field is perpendicular to the current. It is this in-plane anisotropy that leads to the PHE and AMR. Therefore, the observation of PHE, AMR, and negative longitudinal MR are strong evidence of chiral anomaly. In particular, the type-II Dirac and Weyl semimetals should show chiral anomaly only when the magnetic field is within the tilted cone.  Otherwise, the Landau-level spectrum would be gapped without chiral zero mode~\cite{Udagawa2016Chiral, Soluyanov2015WeylII}. This has been supported experimentally in WTe$_2$~\cite{Li2017WTe2, Li2019WTe2}. Therefore, the field-angle dependent chiral anomaly in type-II Dirac and Weyl semimetals is predicted to have switching effect on phenomena such as magnetotransport and magneto-optical conductivity~\cite{Lv2021Topo, Armitage2018Topo, Udagawa2016Chiral}.
\begin{figure}[!tb]\centering
      \resizebox{9cm}{!}{
              \includegraphics{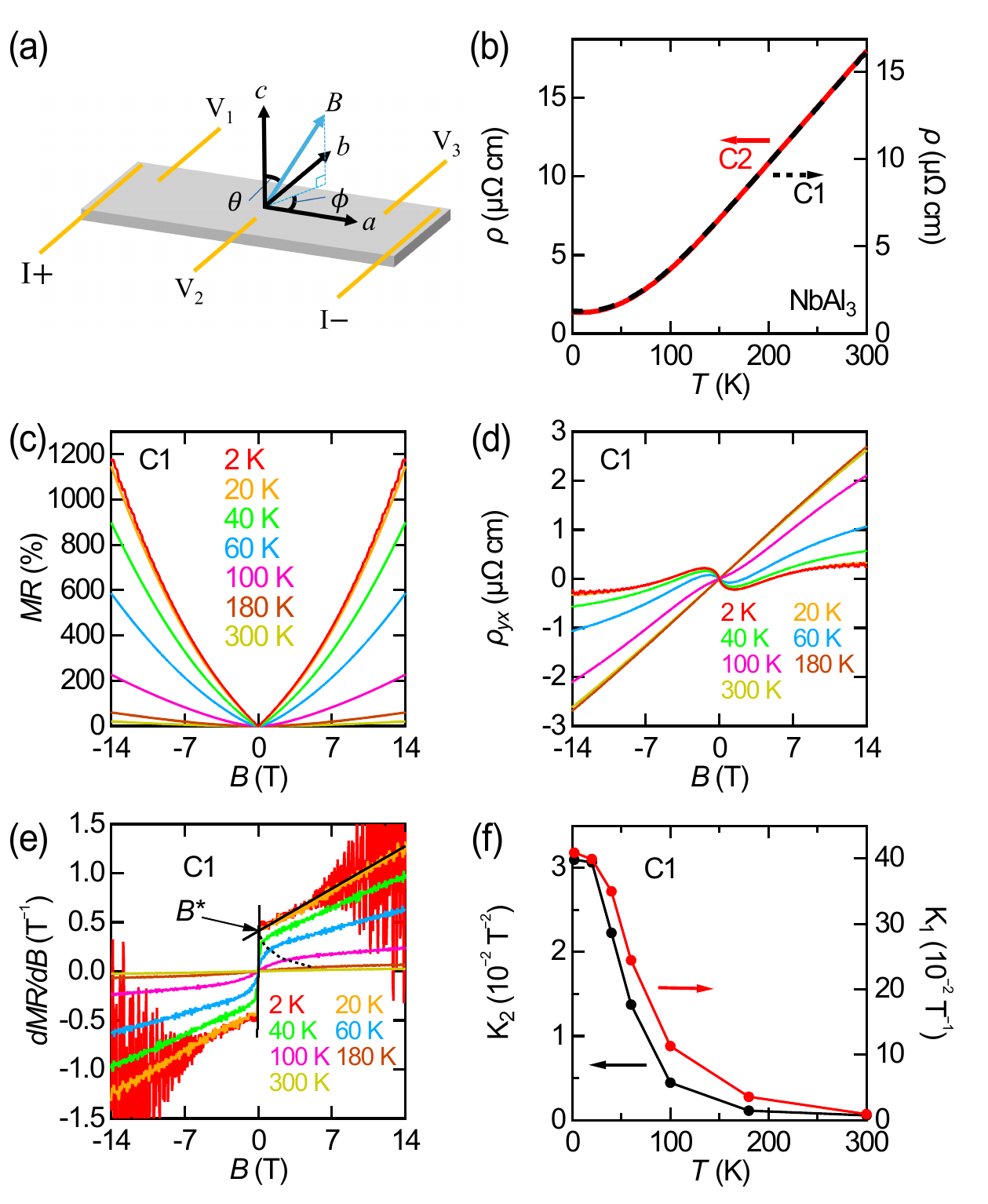}}		
              \caption{\label{Fig1}
(a)~Schematic for the magnetotransport measurement setup. (b)~The temperature dependence of the longitudinal resistivity of the two NbAl$_3$ samples, C1 and C2. Field dependence of (c) the transverse MR and (d) the Hall resistivity at various temperatures, collected on C1. (e) The first derivative of the transverse MR against field at various temperatures. The black solid lines show the definition of $B\text{*}$ for the 2~K dataset. The black dotted line traced out the $B\text{*}$ at different temperatures. (f) The fitted coefficients $K_1$ and $K_2$.}
\end{figure}

Other interesting aspects of the Dirac and Weyl semimetals include the presence of high-mobility carriers and non-trivial Berry phase~\cite{Hu2019Topo}. The presence of high-mobility carriers can be discerned by analyzing the transverse MR and conventional Hall resistivity, measured with an out-of-plane magnetic field as opposed to the in-plane magnetic field for PHE and AMR. The non-trivial Berry phase arises from the cyclotron motion of the carriers encircling Dirac or Weyl points~\cite{Hu2019Topo, Berry1984Berry, Xiao2010Berry, Mikitik1999Berry}. Therefore, transverse MR, conventional Hall resistivity, and Berry phase analysis have been commonly used to probe Dirac and Weyl semimetals, see e.g. Refs.~\cite{Chen2018MAl3, Singha2018VAl3, Hu2016TaP, Huang2015TaAs, Li2019WTe2, Fei2017PdTe2, Li2017WTe2, Chen2016MoTe2, Liang2015Cd3As2}.

Recently, the type-II Dirac semimetallic states in transition-metal icosagenides, MA$_3$ (M$=$V, Nb, Ta; A$=$Al, Ga, In), have been identified with band structure calculations~\cite{Chang2017MA3, Ge2017MA3, Chen2018MAl3}. In particular, the calculated band structures of VAl$_3$, NbAl$_3$, and TaAl$_3$ have been supported by de Haas–van Alphen~(dHvA) oscillations~\cite{Chen2018MAl3}. Among them, VAl$_3$ has the Dirac node closest to the Fermi energy~\cite{Chang2017MA3, Ge2017MA3, Chen2018MAl3}. Therefore, it has attracted the most attention and interesting transport phenomena such as the PHE and AMR ~\cite{Singha2018VAl3, Liu2020VAl3}, and the non-saturating non-quadratic MR~\cite{Singha2018VAl3} have been observed. With the type-II Dirac node further away from the Fermi energy, NbAl$_3$ receives less attentions. Magnetotransport studies have been lacking. 

In this manuscript, we first describe the magnetotransport in NbAl$_3$ in out-of-plane magnetic field ($B\parallelsum c$). The detection of the linear component in the transverse MR and the emergence of high-mobility carriers at low temperatures revealed by the mobility spectrum analysis (MSA) support the quantum limit picture, {\it i.e.} when one or several bands are in the lowest Landau level. Such results cannot be explained by the type-II Dirac node, which is well above the Fermi energy. We then report the magnetotransport in in-plane magnetic field ($B\perp c$) which is perpendicular to the tilting direction of the type-II Dirac cone. Remarkably, PHE, AMR, and negative longitudinal MR, which together are a strong indication of chiral anomaly, have been observed. These data support the existence of additional Dirac or Weyl nodes. Finally, we compare the angular dependence of Shubnikov–de Haas~(SdH) oscillations with our calculated band structure. These confirm the type-II Dirac node is indeed far above the Fermi energy and facilitate the identification of a type-I Dirac node, justifying the magnetotransport in NbAl$_3$.

High-quality NbAl$_3$ crystals have been grown using the flux method. X-ray diffraction on a powdered sample, displayed in Fig.~S1~\cite{Supp}, shows the single-phase nature. Electrical contacts have been made on the crystals with gold wires and silver paste. Two single crystals were used: C1 has dimensions ($l\times w\times t$) approximately 410~$\mu$m~$\times$~200~$\mu$m$~\times$~5~$\mu$m and C2 has dimensions approximately 270~$\mu$m~$\times$~250~$\mu$m~$\times$~5~$\mu$m. Thin samples were used to boost the resistive signals and to eliminate the possibility of current jetting, which can also lead to a negative MR~\cite{Dos2016Jetting}. To avoid current jetting, the current contacts were also made as wide as possible, covering the whole width of the samples. Magnetic field up to 14~T at different field angles and temperature down to 2~K have been achieved with Physical Property Measurement System~(PPMS) by Quantum Design. For temperatures down to 12~mK, the Bluefors Dilution Fridge has been used. To extract the information about the carrier mobilities, mobility spectrum analysis~(MSA)~\cite{Beck1987Msa, Antoszewski1995Msa, Vurgaftman1998Msa, Kiatgamolchai2002Msa, Rothman2006Msa, Beck2021Msa} has been performed on the transverse MR and conventional Hall resistivity data obtained with magnetic field along the $c$-axis. In this work, we have employed the maximum entropy approach introduced by Kiatgamolchai~\etal~\cite{Kiatgamolchai2002Msa}. To ensure optimal fittings to $\sigma_{xx}$ and $\sigma_{xy}$, data of both field orientations (\ie parallel and anti-parallel to $c$) have been  analysed.

Figure~\ref{Fig1}(b) shows the temperature dependence of the resistivity, $\rho(T)$, for two NbAl$_3$ samples. The two samples show only a small difference in the resistivity, attributable to the uncertainty in determining sample dimensions, with C1 having a smaller overall resistivity. However, the two datasets can be scaled perfectly onto each other, resulting in an identical residual resistance ratio [$RRR\equiv R(300~\text{K}/R(2~\text{K})$] of $\sim13$. Additionally, our magnetotransport data from both samples with $B\parallelsum c$ give the same conclusions, as discussed in Ref.~\cite{Supp}. Thus, the two samples are practically the same, and we do not distinguish between them henceforth. The resistivity decreases monotonically from 300~K and forms a plateau at low temperatures down to 12~mK, exhibiting a typical metallic behaviour. In contrast to a previous report of superconductivity at 64~mK~\cite{Leyarovski1977NbAl3}, our resistivity data show no sign of superconductivity.
 
NbAl$_3$ has a body-centered tetragonal lattice structure. Applying an out-of-plane magnetic field along the $c$-axis~[see Fig.~\ref{Fig1}(a)], we measure the transverse MR and conventional Hall effect. Figure~\ref{Fig1}(c) shows the field dependence of the transverse MR at different temperatures. All the transverse MR curves are non-saturating at 14~T. At 2~K, 14~T, the transverse MR reaches 1180~$\%$. Even before subtracting the background, the SdH oscillations are visible in the 2~K data.

To examine the curvature of the transverse MR, we plot $d({\rm MR})/dB$ against the field as shown in Fig.~\ref{Fig1}(e). The strong oscillations in the 2~K dataset are due to the SdH effect. At 300~K, $d({\rm MR})/dB$ versus $B$ has zero intercept and is almost linear, indicating that the transverse MR is approximately quadratic for the entire $B$ range. However, at lower temperatures, the transverse MR is only quadratic up to a limited field approximated by a characteristic field $B\text{*}$. At higher field, $d({\rm MR})/dB$ saturates to a reduced slope. Importantly, linear extrapolation of $d({\rm MR})/dB$ from the high field results in a non-zero intercept, indicating a linear-in-$B$ contribution. Here, $B\text{*}$ is determined to be the field at which the linear extrapolation of $d({\rm MR})/dB$ from the low field and that from the high field intersect, as exemplified by the fittings with the 2~K dataset in Fig.~\ref{Fig1}(e). The $B\text{*}$ for the various temperatures is traced out by the black dotted line in Fig.~\ref{Fig1}(e). Thus, the transverse MR above $B\text{*}$ can be expressed as ${\rm MR}=K_{1}B+K_{2}B^2$, where $K_1$ and $K_2$ are the coefficients, and their temperature dependencies are depicted in Fig.~\ref{Fig1}(f). As the temperature decreases, both $K_1$ and $K_2$ increase in a similar fashion. The linear-in-$B$ MR contribution can be understood if one or some of the bands enters the quantum limit, and if these bands dominate magnetotransport ~\cite{Abrikosov1998Mr, Abrikosov2000Mr, Niu2017Mr}. To be in the lowest Landau level at $T$, we require $k_BT<\Delta_{LL}$, where $\Delta_{LL}$ is the separation between the lowest Landau level and the adjacent Landau level. The separation depends on the details of the band dispersion, but in general $\Delta_{LL}$ is larger for higher field. For parabolic bands, the field needed to satisfy such requirement is very large. However, for the Dirac state, the field needed is much smaller. Therefore, our observation of the linear MR contribution that sets in at a rather low $B$ is possibly due to a small number of Dirac fermions. Yet, these fermions cannot be related to the type-II Dirac node in NbAl$_3$ because the quantum limit picture requires the node to be close to the Fermi energy. The type-II Dirac node in NbAl$_3$ is predicted to be far above the Fermi energy~($\sim230~\text{meV}$). Thus, this leads to the possibility of additional Dirac or Weyl nodes in NbAl$_3$ that has been overlooked.

Figure~\ref{Fig1}(d) displays the field dependence of the Hall resistivity at different temperatures. At 300~K and 180~K, the Hall resistivity is linear in $B$ and the slope is positive. As the temperature decreases from 180~K, the Hall resistivity slope decreases and becomes non-linear in $B$. Below 100~K, the data shows a negative slope at a low field. Since high-mobility carriers have a stronger effect on the low-field magnetotransport, the low-field negative Hall resistivity slope at low temperatures hints at the emergence of the high-mobility electrons. Overall, the Hall resistivity implies the multi-carrier transport with both holes and electrons in NbAl$_3$.
\begin{figure}[!tb]\centering
      \resizebox{9cm}{!}{
              \includegraphics{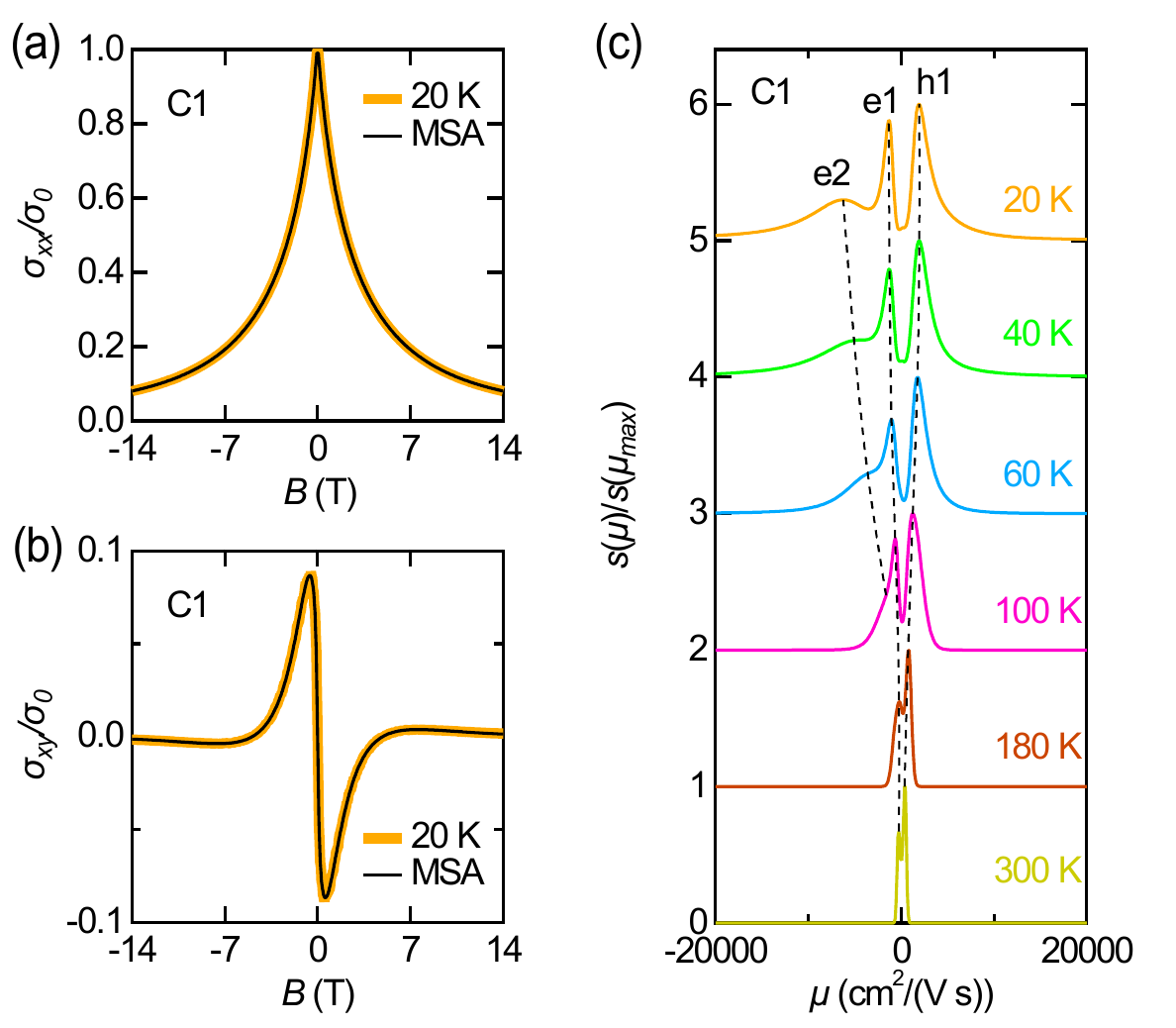}}		
              \caption{\label{Fig2}  
Mobility spectrum analysis~(MSA) for sample~C1. Field dependence of the conductivity tensor component (a)~$\sigma_{xx}$ and (b)~$\sigma_{xy}$ at 20~K fitted by MSA. (c)The mobility spectra normalized to the maximum values $s(\mu_{max})$ at various temperatures.}
\end{figure}

To extract the carrier mobilities, the transverse MR and the conventional Hall resistivity data have been examined simultaneously with MSA, which provides an unbiased interpretation of the magnetotransport without assuming the number of carrier types~\cite{Perdew1996, Koepernik1999, Farrar2022FeSe, Huynh2014FeSe, Huynh2014FeSe}. Figures~\ref{Fig2}(a) and \ref{Fig2}(b) show the MSA fittings to the $\sigma_{xx}$ and $\sigma_{xy}$ measured at 20~K. The fittings to the higher-temperature data, and the mathematical framework behind the MSA procedure, are shown in the Supplemental Material~\cite{Supp}. Figure~\ref{Fig2}(c) shows the resultant mobility spectra at different temperatures. At 300~K, two mobility peaks are resolved: one hole peak with mobility $380~\text{cm}^{2}\text{V}^{-1}\text{s}^{-1}$, and one electron peak with mobility $240~\text{cm}^{2}\text{V}^{-1}\text{s}^{-1}$. As temperature decreases, the mobility corresponding to each peak increases. Also, a high-mobility electron peak emerges and becomes completely developed at 20~K. At 20~K, the two electron peaks have mobilities of $6300~\text{cm}^{2}\text{V}^{-1}\text{s}^{-1}$ and $1400~\text{cm}^{2}\text{V}^{-1}\text{s}^{-1}$, while the hole peak has mobility of $2200~\text{cm}^{2}\text{V}^{-1}\text{s}^{-1}$. Thus, the MSA has shown that the low temperature magnetotransport involve one type of holes and two types of electrons, with one type of electrons having a much higher mobility. By performing further analysis on the partial conductivity due to electrons with a multicarrier model, we estimate that the carrier density of the Dirac electrons at 20~K is $\sim3.51\times10^{19}$~cm$^{-3}$.  For a comparison, the same analysis gives the carrier density of `normal' electrons of $7.51\times10^{20}$~cm$^{-3}$. The small density of Dirac electrons translates to a quantum oscillation frequency of only $\sim$15~T when $B\parallelsum c$, when the multiplicity of the associated Fermi surface is considered~\cite{Supp}.

\begin{figure}[!tb]\centering
      \resizebox{8.5cm}{!}{
              \includegraphics{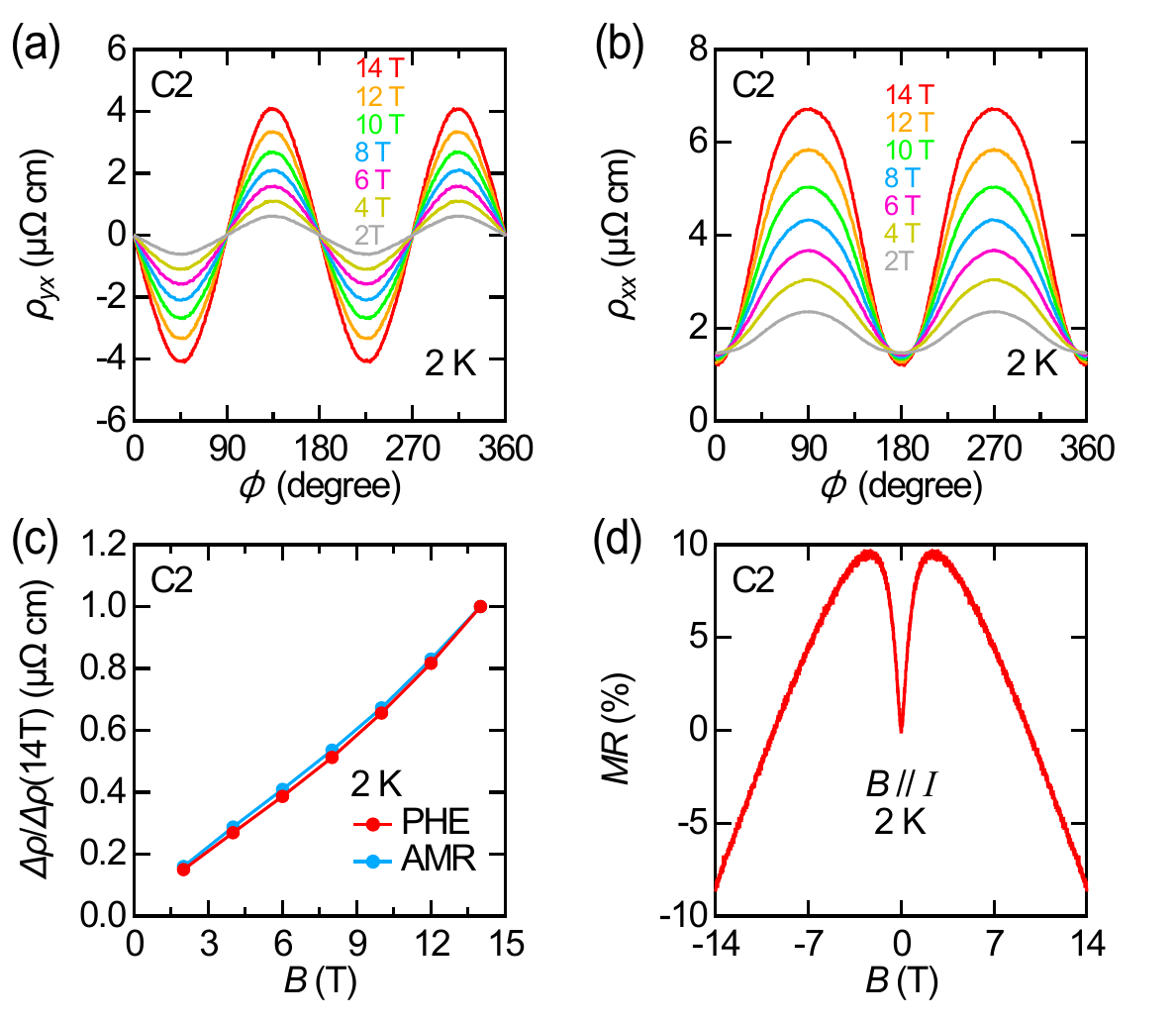}}		
              \caption{\label{Fig3}
The (a)~PHE and (b)~AMR at different magnetic fields at 2~K. (c)~The field dependence of the resistivity anisotropy~$\Delta\rho$ from the PHE data and that from the AMR data. (d)~The longitudinal magnetoresistance at $\phi=0^\circ$ and 2~K.}
\end{figure} 
Next, we present the in-plane magnetotransport data against the field angle~$\phi$ while keeping $\theta=90^\circ$, where $\phi$ and $\theta$ are defined in Fig.~\ref{Fig1}(a). Surprisingly, clear PHE and AMR signals are observed, as shown in Figs.~\ref{Fig3}(a) and \ref{Fig3}(b). PHE and AMR arise from a resistivity anisotropy $\Delta\rho\equiv\rho_{\perp}-\rho_{\parallel}$, where $\rho_{\perp}$ and $\rho_{\parallel}$ correspond to the longitudinal resistivity when the field is perpendicular and parallel to the current, respectively. The resistivity tensor components associated with the PHE and AMR follow the relations~\cite{Zhong2023Phe}
\begin{subequations}
    \begin{align}
        &\rho^{\text{PHE}}_{yx}=-\Delta\rho\sin\phi\cos\phi~\text{ and}\label{PheEqtA}\\
        &\rho^{\text{AMR}}_{xx}=\rho_\perp-\Delta\rho\cos^2\phi~\text{.}\label{PheEqtB}
    \end{align}
\end{subequations}
Our data can be well-fitted to Eqns.~(\ref{PheEqtA}) and (\ref{PheEqtB}). Figure~\ref{Fig3}(c) shows the field dependence of the normalized resistivity anisotropy~$\Delta\rho/\Delta\rho(14~\text{T})$ obtained from the fittings. Both the PHE and AMR show consistent normalized $\Delta\rho$ that increases with the field, confirming the common origin. As shown in Fig.~\ref{Fig3}(b), when the applied field is parallel to the applied current, \ie at $\phi=0^\circ$, 180$^\circ$, and 360$^\circ$, the resistivity can be lower at higher magnetic fields. This implies negative longitudinal MR, which we have measured at 2~K against the magnetic field applied at $\phi=0^\circ$. The result is displayed in Fig.~\ref{Fig3}(d). For the chiral anomaly, $\rho_{\perp}$ increases while $\rho_{\parallel}$ decreases with an increasing magnetic field~\cite{Liang2018Phe}. These lead to a $\Delta\rho$ that increases with an increasing magnetic field. Thus, our in-plane magnetotransport result is consistent with the chiral anomaly effect. According to Chang~\etal~\cite{Chang2017MA3}, the tilting direction of the type-II Dirac cone is the $k_z$ direction, which is parallel to $c$. By applying a magnetic field in the $a$-$b$ plane (where $a=b$), the possible chiral anomaly due to the type-II Dirac node~\cite{Chang2017MA3, Ge2017MA3, Chen2018MAl3} is excluded. Thus, our data provide evidence of additional Dirac or Weyl nodes, which can either be type-I or type-II, in NbAl$_3$. For the case of type-II, the tilting direction of the corresponding cones would be of certain range from the $a$-axis. 

The clear SdH oscillations in the transverse MR at low temperatures provide an avenue to compare the band structure with previously published results. SdH oscillations are obtained by removing the MR background via low-order polynomial fittings, as exemplified by the 2~K dataset shown in the inset of Fig.~\ref{Fig4}(a). For oscillatory signals at higher temperatures, see Supplemental Material~\cite{Supp}. The main panel of Fig.~\ref{Fig4}(a) shows the fast Fourier transform~(FFT) of the SdH oscillations for $B\parallelsum c$ at different temperatures. Six peaks have been identified, namely $\beta_1$ at $\sim$190~T, $2\beta_1$ at $\sim$379~T, $3\beta_1$ at $\sim$570~T, $\beta_2$ at $\sim$507~T, $2\beta_2$ at $\sim$1010~T, and $\beta_1+\beta_2$ at $\sim$691~T. $\beta_{1}$ and $\beta_2$ correspond respectively to the “neck” and the “belly” of the dumbbell-like hole pocket (See Fig. S8 in the Supplemental Material~\cite{Supp}). The fact that our SdH data are consistent with the de Haas-van Alphen data collected via torque magnetometry~\cite{Chen2018MAl3} implies the detection of the bulk electronic structure. We did not detect the electron pocket, which is also consistent with Chen \etal's report that its detection requires a much higher field~\cite{Chen2018MAl3}. Fitting the temperature dependence of the peak amplitude to the thermal damping factor [$R_T=14.69(m\text{*}T/B)/\sinh{(14.69m\text{*}T/B)}$] of the Lifshitz-Kosevich~(LK) theory gives the effective mass~$m\text{*}$. Figure~\ref{Fig4}(b) displays the results for the $\beta_1$, $2\beta_1$, $3\beta_1$, and $\beta_2$ peaks. The corresponding effective masses are $\sim$0.20~$m_e$, $\sim$0.39~$m_e$, $\sim$0.53~$m_e$, and $\sim$~0.51~$m_e$ respectively, where $m_e$ is the rest mass of an electron. The effective masses corresponding to $2\beta_1$ and $3\beta_1$ are approximately integer multiples of those of $\beta_1$, as expected for harmonics.

\begin{figure}[!tb]\centering
      \resizebox{8.5cm}{!}{
              \includegraphics{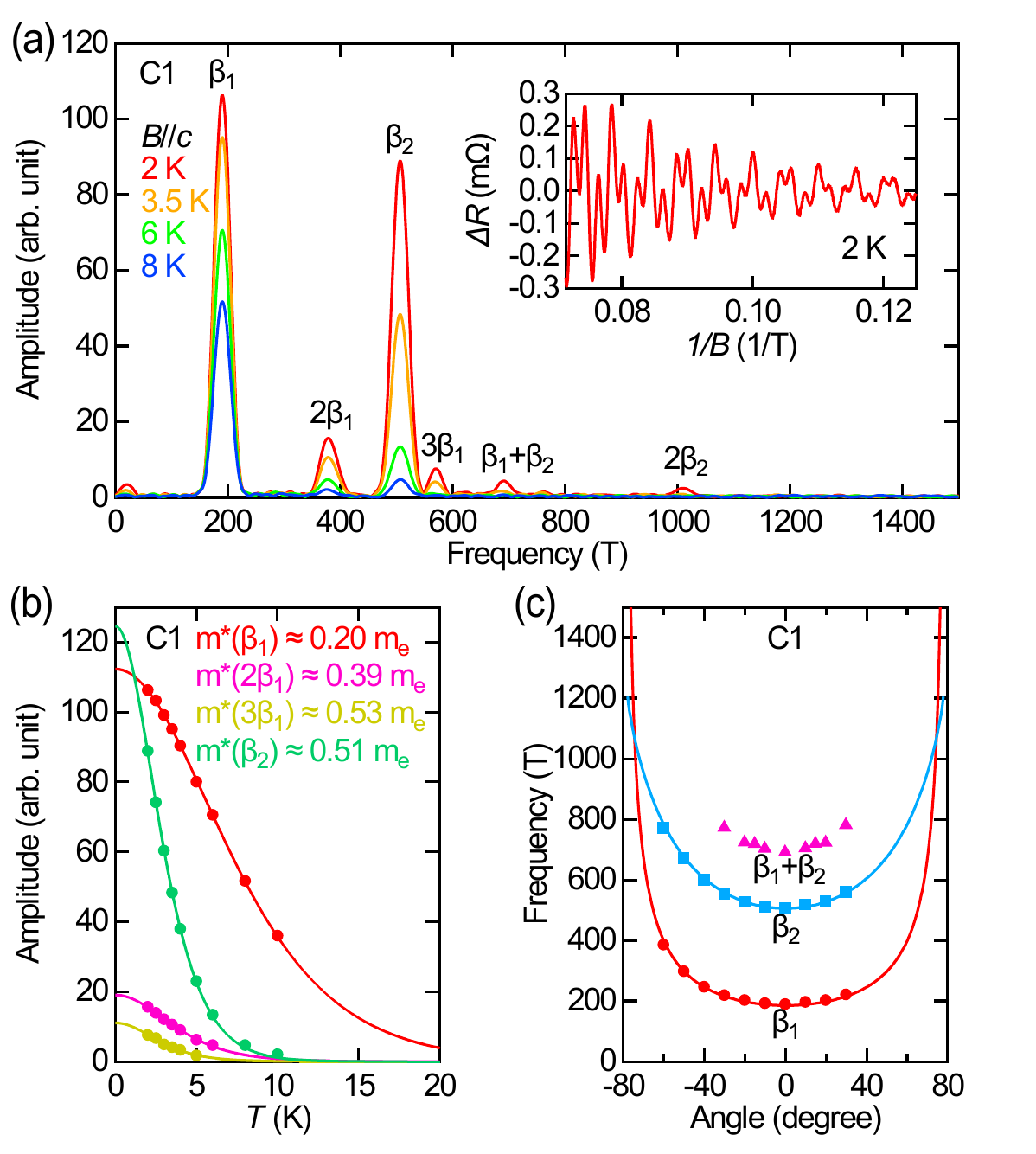}}		
              \caption{\label{Fig4}  
SdH oscillations from sample~C1. (a) FFT spectra at different temperatures. The inset shows the oscillatory signals at 2~K after removing the magnetoresistance background. (b) Temperature dependence of the oscillation amplitudes of the $\beta_1$, $2\beta_1$, $3\beta_1$ and $\beta_2$ peaks. The solid curves are the fittings using the thermal damping factor~($R_T$) of the Lifshitz-Kosevich theory. (c) Angular dependence of the oscillation frequencies. The symbols represent the experimental data. The lines represent the extracted $\beta_1$ and $\beta_2$ frequencies from the calculated band structure with suitable scaling factors (see text).}
\end{figure} 

Figure~\ref{Fig4}(c) displays the angular dependence of the frequencies, collected by varying $\theta$ while keeping $\phi=0^\circ$. For clarity, the harmonics are not shown in Fig.~\ref{Fig4}(c). The frequency of $\beta_1+\beta_2$ equal to the sum of the frequencies of $\beta_1$ and $\beta_2$ at all angles. Our band structure calculation shows the type-II Dirac node $\sim$220~meV above the Fermi energy along the $\Gamma$--M direction, consistent with previous works~\cite{Chang2017MA3, Ge2017MA3, Chen2018MAl3}. Interestingly, a type-I Dirac node $\sim$60~meV below the Fermi energy can be identified in the dispersion along the M--Y$_1$ direction. Detailed calculation results can be found in~\cite{Supp}. We show in Fig.~\ref{Fig4}(c) the quantum oscillation frequencies extracted from the band structure. With scale factors of 1.8 for $\beta_1$ and 1.15 for $\beta_2$, we can bring our data to a good agreement with the calculation. The need for these factors can be understood as the discrepancy between the warping of the Fermi surface deduced from the experiment and that from the calculation. Such discrepancy may stem from the slight difference in the lattice constant $c$. $\beta_1$ corresponds to the hole-like band crossing the Fermi level along the $\Gamma$--X direction in the band structure. To have a SdH frequency that is 1.8 times larger, the band can be shifted upward relative to the Fermi energy by $\sim$55~meV. Assuming all other bands can be shifted by the same amount, the type-I Dirac node along the M--Y$_1$ direction would be less than $\sim$10~meV from the Fermi energy, justifying our experimental results. 

In summary, we have demonstrated the plausible existence of additional Dirac or Weyl nodes in NbAl$_3$. Our measurements with out-of-plane magnetic field reveal the linear-in-$B$ component in the transverse MR and the emergence of high-mobility electrons at low temperatures. This leads to the possibility of Dirac or Weyl nodes close to the Fermi energy, in addition to the previously predicted type-II Dirac node that is far above the Fermi energy. Furthermore, we conduct measurements in in-plane magnetic field, in which we detect compelling evidence of chiral anomaly through the observed planar Hall effect, anisotropic MR and negative MR. However, this magnetic field direction is perpendicular to the tilting direction of the type-II Dirac cone. Therefore, the detected chiral anomaly is not related to the predicted type-II Dirac node. Our Shubnikov-de Haas data further agree with the Fermi surface predicted by the same calculations that reported the type-II Dirac node. Overall, the magnetotransport phenomena we uncovered imply the existence of additional Dirac or Weyl nodes in NbAl$_3$ that may have previously been overlooked.

\begin{acknowledgments}
\section{Acknowledgements}
The work was supported by Research Grants Council of Hong Kong (A-CUHK 402/19, CUHK 14301020, CUHK 14300722), CUHK Direct Grant (4053577, 4053525) JSPS KAKENHI (JP22H01181, JP23K19032, JP22H04933, JP23K22452, JP23K17353, JP24H00007), and National Science and Technology Council of Taiwan (112-2124-M-006-009 and 113-2112-M-006-009-MY2). 
\end{acknowledgments}


\providecommand{\noopsort}[1]{}\providecommand{\singleletter}[1]{#1}%

\end{document}